# An Intrinsic Entropy Model for Exchange-Traded Securities

**Claudiu Vințe [1,*], Ion Smeureanu [1], Titus-Felix Furtună [1] and Marcel Ausloos [2,3]**

[1] Department of Economic Informatics and Cybernetics, Bucharest University of Economic Studies, 15–17 Calea Dorobanților, 010552 Bucharest, Romania; ion.smeureanu@ie.ase.ro (I.S.); felix.furtuna@ie.ase.ro (T.-F.F.)

[2] School of Business, Brookfield, University of Leicester, Leicester LE2 7RQ, UK; marcel.ausloos@uliege.be

[3] GRAPES, 483 rue de la Belle Jardiniere, B-4031 Liege, Belgium

[*] Correspondence: claudiu.vinte@ie.ase.ro; Tel.: +40-751-251-119

**Abstract:** This paper introduces an intrinsic entropy model which can be employed as an indicator for gauging investors' interest in a given exchange-traded security, along with the state of the overall market corroborated by individual security trading data. Although the syntagma of intrinsic entropy might sound somehow pleonastic, since entropy itself characterizes the fundamentals of a system, we would like to make a clear distinction between entropy models based on the values that a random variable may take, and the model that we propose, which employs actual stock exchange trading data. The model that we propose for the intrinsic entropy does not include any exogenous factor that could influence the level of entropy. The intrinsic entropy signals if the market is either inclined to buy the security or rather to sell it. We further explore the usage of the intrinsic entropy model for algorithmic trading, in order to demonstrate the value of our model in assisting investors' intraday stock portfolio selection, along with timely generated signals for supporting the buy/sell decision-making process. The test results provide empirical evidence that the proposed intrinsic entropy model can be used as an indicator for evaluating the direction and the intensity of intraday trading activity of an exchange-traded security. The data employed for testing consisted of historical intraday transactions executed on The Bucharest Stock Exchange (BVB).

## 1. Introduction and Context

Investing in stocks or any other exchange-traded securities is a decision-making process under risk. Therefore, investors may have some knowledge regarding the probability of price rise or fall for a given stock. This knowledge is achieved through data analysis, taking into account information available to investors from various sources:

- global, regional or national reports regarding the general state of the economic environment;
- company financial reporting, the basis of fundamental analysis for any exchange-listed company;
- specific market news regarding a certain economic sector, activity or company [1,2];
- stock exchange transaction data, disseminated to investors mainly through electronic data feeds and consisting in prices, volumes, timestamps of executed trades [3].

In this paper we investigate how the latter source of information, the trade data, can offer itself an insightful perspective upon investors' interest in buying or selling a listed stock. Historical data shows us that stock prices cannot be consistently predicted for a long period of time, whether we consider Fama's efficient-market hypothesis (EMH) being at work or not [4]. For those involved in trading activity, either as investors or as specialists in algorithmic trading, the ability to react promptly to market changes is critical. For risk mitigation purpose, such ability ranks higher in investors' priorities than striving to discover patterns that may potentially lead to anticipate market changes. In order to assist investors with such ability, of reacting promptly to market evolution, we propose in this paper an intrinsic entropy model for exchange-traded securities, based on intraday trade data generated by the stock exchange. In addition, we used this intrinsic entropy model for algorithmic trading, in order to demonstrate the value of our model in assisting investors' intraday stock portfolio selection, along with timely generated signals for supporting the buy/sell decision-making process. The intrinsic entropy signals if the market is either inclined to buy the security or rather to sell it. The test results provide empirical evidence that the proposed intrinsic entropy model can be used as an indicator for evaluating the direction and the intensity of intraday trading activity of an exchange-traded security. Our intrinsic entropy model starts from the observation that if we divide the quantity executed through each individual transaction ($q_i$) to the cumulated executed quantity ($Q$) up to given moment in time ($t$) during the trading session:

$$Q = \sum_{i=1}^{t} q_i \qquad (1)$$

We obtain fractions $\frac{q_i}{Q}$ that may play the role of probabilities in the standard entropy model. If we take into account also the prices at which the trades are made, then these probabilities could be considered as degrees of confidence that investors give to each price level.

Although the concept of entropy originated in physics, it has been constantly extended to and employed in numerous other disciplines, ranging from statistical mechanics and information theory, to economics and biology. However, there is neither a unique nor a unified interpretation of the entropy concept. Even within the same domain, entropy can be defined differently, depending on different contexts. Most of the entropy models are defined in terms of probabilities. Nevertheless, even when entropy is delineated in terms of probabilities, these could be either interpreted as chances of occurrence for a physical phenomenon, or credences, as degrees of belief provided by the subjects of a sociological sample [5].

Since its inception, Shannon's [6] seminal concept of information entropy has known multiple interpretations and has seen various extensions. Naming it after Boltzmann's H-theorem [7,8], Shannon denoted the entropy *H* of a discrete random variable *X* with possible values *{x₁, x₂, …, xₙ}* as:

$$H(X) = E(I(X)) \qquad (2)$$

where *E* is the expected value and *I* is the information content of *X*.

*I(X)* is itself a random variable. If *p* denotes the probability mass function of *X*, then entropy can explicitly be written as:

$$H(X) = \sum_{i=1}^{n} p(x_i)(I(x_i)) = \sum_{i=1}^{n} p(x_i) log_b \frac{1}{p(x_i)} = - \sum_{i=1}^{n} p(x_i) log_b p(x_i) \qquad (3)$$

where *b* is the base of the logarithm used. Common values of *b* are 2, Euler's number *e*, and 10. Correspondingly, the unit of measure for entropy is *bit* for *b* = 2, *nat* for *b* = e, and *digit* for *b* = 10 [9].

If there are cases where $p_i$ = 0, then the value of the corresponding summand (*0 $log_b$ 0*) becomes zero as well. Applying l'Hôpital's rule to the following limit, the proof can be obtained immediately:

$$\lim_{p \to 0+} p \log p = \lim_{p \to 0+} \frac{\log p}{1/p} = \lim_{p \to 0+} \left( \frac{1/p}{-1/p^2} \right) = \lim_{p \to 0+} (-p) = 0 \qquad (4)$$

It can be shown that, for a given number of system states *n* (meaning that *n* is fixed), along with their associated probability of occurrence, the uncertainty *H* reaches its maximum value only when the states are equally probable (*EP*) [9,10]. 'Equally probable' means that $p_i = 1/n$. Therefore, if we substitute this into the entropy Equation (3), we obtain:

$$H_{EP} = -\sum_{i=1}^{n} \frac{1}{n} log_b \frac{1}{n} \quad (5)$$

Since *n* is not a function of *i*, we can pull it out of the sum:

$$H_{EP} = -\left(\frac{1}{n} log_b \frac{1}{n}\right) \sum_{i=1}^{n} 1 = -\left(\frac{1}{n} log_b \frac{1}{n}\right) n = -log_b \frac{1}{n} = log_b n \quad (6)$$

During the early 1970s, entropy and information theory analysis experienced a brief surge in interest in finance and economics literature [11–14].

Around the middle of 1970s, Horowitz and Horowitz questioned the underlying basis of employing entropy analysis in economics and finance [15]. They attempted to conclude whether entropy analysis measures worthwhile information which wouldn't be otherwise available to standard statistical techniques, such as variance or correlation analysis. Philippatos and Wilson argue in their papers [16,17] that entropy is a better statistical measure of risk than variance, since entropy does not make assumptions regarding the underlying probability distribution. In contrast, White argued that, since entropy analysis is not integrated into economic theory, including the theory of choice under uncertainty, it should not be used in studying economic and financial phenomena [18].

The information entropy proposed by Shannon and its developments in the economics literature generally uses the convention of having *C* for the original Boltzmann constant $k_B$ equal to 1.

$$H = -C \sum_{i=1}^{n} p_i log p_i = -\sum_{i=1}^{n} p_i log p_i \quad (C=1) \quad (7)$$

From this general method of computing the entropy, in 1986 Nawrocki and Harding introduced the concept of weighted entropy [19]. They noted that there appears to be no *a priori* reason why *C* should be constant for all the observed or considered system states.

In 1988 Tsallis introduced the framework for generalizing the standard statistical mechanics [20], and proposed a generalization of the standard Boltzmann–Gibbs entropy [21–23].

In their study of 2002, Maasoumi and Racine inferred that the evidence in favour of conditional predictability of stock returns is non-robust with respect to period of analysis and data frequency [24]. Ausloos and Ivanova proposed in 2003 a model encompassing Tsallis nonextensive statistics and bringing about to evolution equations of the Langevin and Fokker-Planck type [25].

More recently, Zhou et al. reviewed in 2013 the concepts and the principles concerning entropy models, along with their applications in the field of finance, especially in portfolio selection and asset valuation [26–29]. Li et al. employed transfer entropy to study the interbank contagion, exploiting the market information based on stock prices [30]. Cerqueti et al. adopted entropies as measures for evaluating the concentration in the market, namely a joint copula-entropy approach [31]. When it comes to using entropy for measuring diversification and corporate growth, Ceptureanu et al. employed an adapted Jacquemin–Berry entropy index, to study both product and international diversification [32].

The intrinsic entropy model that we introduce here relates to the concept of weighted entropy approach proposed by Nawrocki and Harding. The probabilities are weighted with the price variation of the individual trades, from a chosen reference price. The novelty of our intrinsic entropy is underpinned by the interpretive model that we propose for the probabilities. We consider the probabilities employed in the intrinsic entropy computation as degrees of confidence that investors give to trade price levels.

## 2. The Proposed Intrinsic Entropy Model and Computation Methodology

The purpose of the intrinsic entropy model proposed in this paper is for gauging the investors' interest in some given listed stocks. Furthermore, the intrinsic entropy provides an indication regarding the direction and the intensity of the interest, either in buying or selling the exchange-traded security.

A trading day, for each exchange-listed security, consists in a succession of transactions, generated by the exchange matching engine, when buy and sell orders meet the conditions for being partially or entirely executed [33].

The stock market can be considered as an open system, which has as inputs the buy/sell orders placed by the participants, and as outputs the trade data generated and based on the exchange matching algorithm, as illustrated in Figure 1.

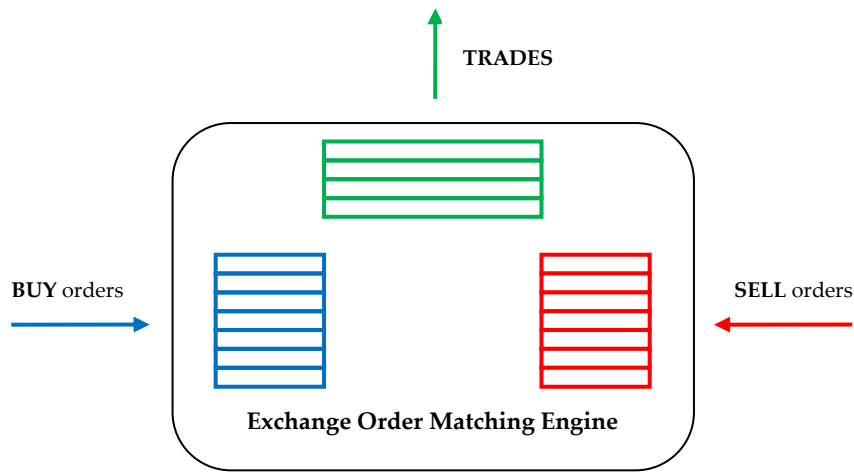

**Figure 1.** Trade generation processes within a stock exchange, based on buy and sell orders placed by investors.

Each individual transaction, namely a trade, consists of the following information:

- the price at which the trade was made;
- the executed quantity, i.e., the quantity that could be put in correspondence by the exchange matching engine from one buy and one sell orders already placed on the market;
- the timestamp at which the order matching occurred and the trade was generated.

Building the intrinsic entropy model starts from the following observations relative to intraday transactions executed on a stock exchange. For any given exchange-listed stock, a trade made on the market at a certain price level is interpreted as follows:

(a) the quantity executed through that transaction, relative to the total quantity executed for the considered stock, up to the moment of the intrinsic entropy calculation $\frac{q_i}{Q}$ represents the degree of confidence or support that the market provides to the price level at which the trade was made;

(b) the price at which the order matching occurs, relative to a certain reference price, offers an indication about the inclination of the investors towards buying or selling the considered stock.

Taking into account these aspects, the intraday intrinsic entropy model that we propose in this paper has the following formalization:

$$H_t^X = -\sum_{i=1}^{n} \left(\frac{p_i}{p_{ref}} - 1\right) \frac{q_i}{Q} \ln\left(\frac{q_i}{Q}\right) \qquad (8)$$

$H_t^X$ being the entropy computed for symbol *X* at moment *t*, and the components are:

- $n$   - total number of trades executed for symbol *X* in the current trading session up to moment *t*
- $i$   - ordinal trade number
- $q_i$   - trade quantity i.e., number of shares of trade *i* for symbol *X*
- $p_i$   - trade price i.e., the price of trade *i* for symbol *X*
- $Q$   - total traded quantity i.e., the number of shares traded during the day for symbol *X*, up to moment *t*

$$Q = \sum_{i=1}^{n} q_i$$

$p_{ref}$ - reference price for symbol *X*, corresponding to the trading data prior to the moment *t*

Regarding the employed reference prices, we experimented with:

(a) the opening price of the current trading day,
(b) the price of the preceding transaction, constructing thereby a Markov chain,
(c) and the volume weighted average price (VWAP), computed with trading data executed over the day up to, and including, the preceding transaction [34].

Operating with the opening price as reference, or any other pegged price for that matter, could generate a decoupling between the evolution of intrinsic entropy and price variation over the day. For example, the stock prices could stay throughout the day below the previous day's closing price, while they may actually go up from a low opening price.

Choosing the volume weighted average price as reference means that, if we consider the trade *n* for the intrinsic entropy computation, then the VWAP is calculated up to, and including the trade *n−1*:

$$p_{ref} = VWAP = \frac{\sum_{i=1}^{n-1} p_i q_i}{\sum_{i=1}^{n-1} q_i} \qquad (9)$$

where $q_i$ and $p_i$ are the quantity and the price, respectively, corresponding to the *i* ordinal trade of the day. Since the entropy is computed every time a new trade is made, then the value for *t* coincides with the ordinal trade number *i*.

The fractions $\frac{q_i}{Q}$ can be assimilated to probabilities, namely the probability of having generated the trade *i* at a certain price level. In addition to this interpretive model, the following condition has to be satisfied:

$$\sum_{i=1}^{n} \frac{q_i}{Q} = 1 \qquad (10)$$

Employing the price of the preceding transaction as reference price preserves the atomicity of each trade within the overall pool of transactions that constitute the trading day on a stock exchange. We are constructing thereby a Markov chain, in which the price of trade is compared only to the price of the preceding trade. Consequently, the price variation weights $\left(\frac{p_i}{p_{n-1}} - 1\right)$ applied to the probabilities $\frac{q_i}{Q}$ reflect local micro changes within the system, and offer the intrinsic entropy model the ability to accrue organically these micro changes, regardless the global indicators, such as the VWAP. Figure 2 illustrates a comparative evolution of the intraday intrinsic entropy, depending of the price chosen as reference: the opening price, the price of the preceding trade and the VWAP. The trade data for the symbol BRD samples of transactions executed throughout the day of December 19, 2018.

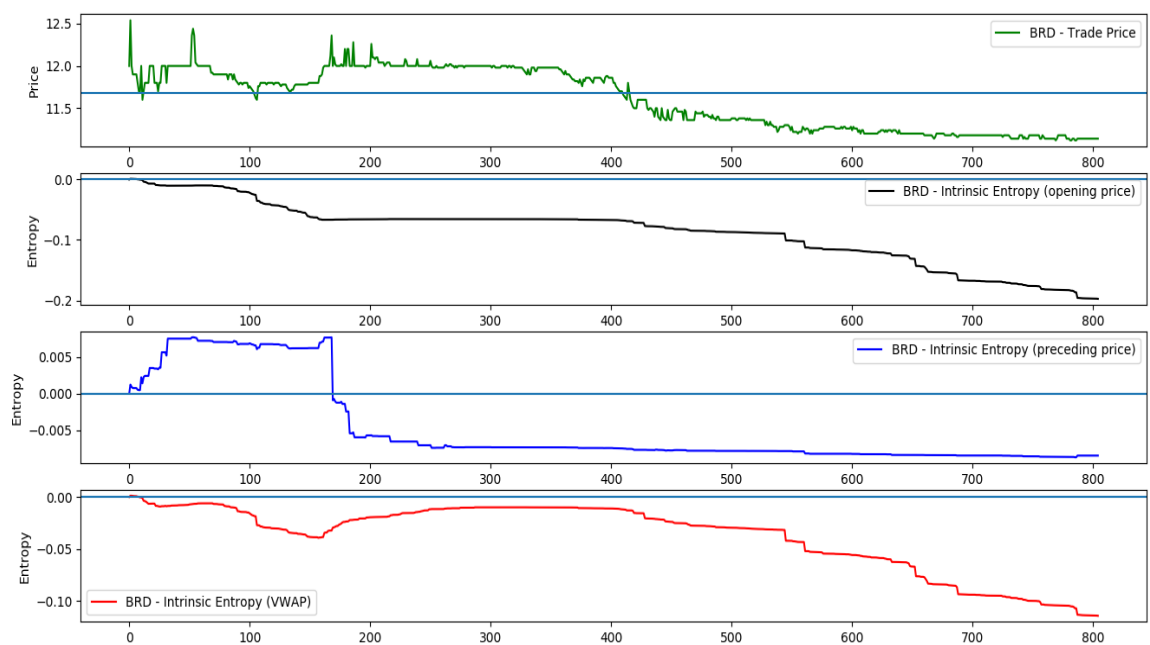

**Figure 2.** A comparative evolution of the intraday intrinsic entropy, depending of the chosen reference price for the day of December 19, 2018 *(*symbol BRD*)*.

Therefore, what particularizes the proposed intraday intrinsic entropy is the fact that it is computed solely based on the actual trading data. The required data for computing $H_t^X$ consists of:

(a) number of trades-the number of generated executions by the exchange matching engine per symbol, during the trading day;
(b) trade executed quantity;
(c) trade executed price;
(d) overall traded volume per symbol, during the day.

From this perspective, we can refer to the $H_t^X$ measure as the intrinsic entropy of a traded stock, or any exchange-traded security: currencies, bonds, exchange-trades funds, exchange-trades notes, etc. Although the syntagma "intrinsic entropy" might sound somehow pleonastic, since the entropy itself characterizes the fundamentals of a system, our intention is to make a clear distinction between entropic models based on the values that a random variable may take, and the model that we propose in this paper, which employs actual stock exchange trading data. There is no explicit exogenous factor to impact the level of entropy in our model. Any external influences are embedded, already contained in the trading data (volume, price, trade frequency, number of transactions), and therefore quantified in the intrinsic entropy.

## 3. Results and Interpretation

In the intrinsic entropy model, the probabilities represented by the fractions $\frac{q_i}{Q}$ (executed quantity of trade *i*, relative to the overall executed quantity during the day, up to the trade *i*), are weighted with the price variation, in terms of the weighted entropy concept introduced by Nawrocki and Harding [19]. The perspective that we propose is slightly different, though. Our interpretive model assimilates these probabilities as degrees of confidence that the market gives to a certain stock price level and, subsequently, to the price variation relative the chosen reference price. Table 1 samples the type of input data that is employed here for computing the intrinsic entropy. It is historical production data, containing intraday trades from December 2018, namely the executed quantity, the executed price, and the timestamp, for a given listed stock on the BVB.

**Table 1.** Trade data for the symbol BRD representing samples of transactions executed during December 19, 2018. Intraday trades executed between 10:00 and 18:00.

| Timestamp | Traded Symbol | Trade Quantity (no of Shares) | Trade Price |
|---|---|---|---|
| 19 December 2018 10:00 | BRD | 57,795 | 12 |
| 19 December 2018 10:00 | BRD | 55 | 12 |
| 19 December 2018 10:00 | BRD | 300 | 12.54 |
| 19 December 2018 10:00 | BRD | 60 | 12 |
| 19 December 2018 10:00 | BRD | 210 | 11.9 |
| 19 December 2018 10:00 | BRD | 220 | 11.9 |
| 19 December 2018 10:00 | BRD | 2 | 11.9 |
| 19 December 2018 10:00 | BRD | 94 | 11.9 |
| … | … | … | … |
| 19 December 2018 17:56 | BRD | 350 | 11.14 |
| 19 December 2018 17:56 | BRD | 60 | 11.14 |
| 19 December 2018 17:56 | BRD | 1000 | 11.14 |
| 19 December 2018 17:56 | BRD | 100 | 11.14 |
| 19 December 2018 17:57 | BRD | 448 | 11.14 |
| 19 December 2018 17:57 | BRD | 1000 | 11.14 |
| 19 December 2018 17:58 | BRD | 1 | 11.14 |

Based on the sampled data, Figure 3 shows the intraday evolution of the entropy for the BVB-listed symbol BRD (BRD – Groupe Société Générale), during 19th of December 2018, along with the evolution of the stock price, traded volume and the values of the entropy components. It has to be noted that the absolute values for the boundaries of the intrinsic entropy are still given by (2) and (3).

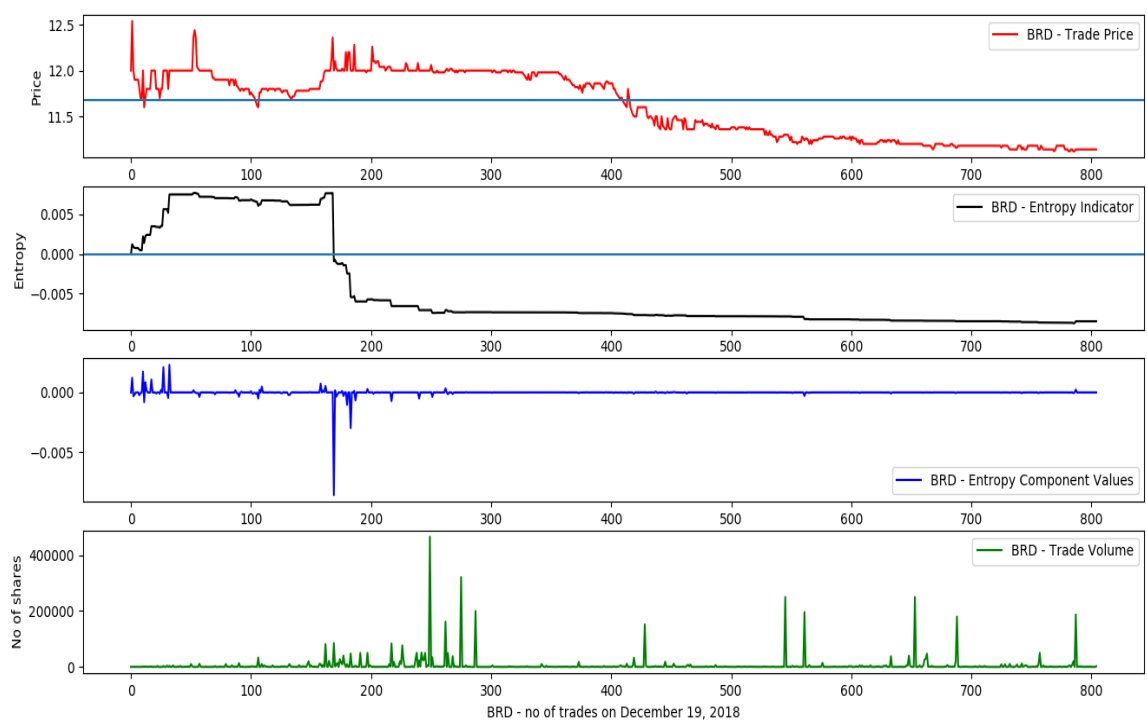

**Figure 3.** Price variation, entropy, entropy component values and trade volumes for the day of December 19, 2018 (symbol BRD*)*.

In order to avoid the impact of the price level on the values of the intrinsic entropy $H_t^X$ we opted for the relative price variation, rather than the absolute price variation from the reference. This approach makes the price weighting component of the intrinsic entropy independent from the individual price level of a traded instrument or another.

One of the salient characteristics of the intrinsic entropy that we introduce in this paper is that it may have either negative or positive values. The absolute value of the intrinsic entropy measures the level of uncertainty that investors assign to an exchange-traded stock. In addition to that, the fact that this intrinsic entropy may have positive or negative values it is meaningful as well. Using historical production intraday executions, our empirical tests show that having the intrinsic entropy entering in the negative territory signifies an inclination of the market towards selling the considered instrument. Reciprocally, the intrinsic entropy has positive values when the market shows a preference for buying the considered exchange-traded instrument.

We consider the intrinsic entropy as an indicator for anticipating the market interest in a given traded stock, interest that can make stock prices go above or below the VWAP.

The absolute value of the intrinsic entropy is explained by the following two underlying phenomena:

(a) the entropy components keep accruing during the day, when execution prices move consistently in a certain direction, up or down;
(b) a high number of transactions, when the stock is heavily traded, suggests that the market, is keenly interested in the security, and this high number of states, and their associated probabilities, support the direction of price evolution.

Regarding the reference price, we determined empirically that a moving reference price reflects more accurately the price weights in the intrinsic entropy model, as is illustrated in Figure 4. The test data consists of trades executed on the BVB for the symbol TGN (Transgaz) throughout the day of December 19, 2018. The price of the preceding transaction turned out to provide the best local indication of the price direction change, with each executed trade, or considered state, in terms of Gibbs entropy.

Relative price variation $\left(\frac{p_i}{p_{n-1}} - 1\right)$ provides both:

(a) anchoring to the probability provided by the fraction $\frac{q_i}{Q}$ ;
(b) indication regarding the direction for the trading activity of a given stock, up to the point in time when the entropy is computed.

If it had not been for the price variation weight, the sum $-\sum_{i=1}^{n} \frac{q_i}{Q} \ln\left(\frac{q_i}{Q}\right)$ would have been positive.

The price variation weight $\left(\frac{p_i}{p_{n-1}} - 1\right)$ has the following implications on the intrinsic entropy, leaving aside the probability $\frac{q_i}{Q}$ associated to the price change occurrence:

(a) if, predominantly, $\left(\frac{p_i}{p_{n-1}} - 1\right) > 0$ then there is a high probability that the computed intrinsic entropy is positive, $H_t^X > 0$ ;
(b) if, predominantly, $\left(\frac{p_i}{p_{n-1}} - 1\right) < 0$ then there is a high probability that the computed intrinsic entropy is negative, $H_t^X < 0$ .

The absolute value of the intrinsic entropy is an indication of the interest that the investors manifest, relative to security *X*, up to the moment *t*. A high absolute value of the intrinsic entropy suggests a great market interest in the given security, either in buying or selling it. Conversely, a low absolute value of the stock intrinsic entropy, a value remaining around zero, indicates a low interest showed by the investors in the considered security. Small or non-existent price variation means that $\left(\frac{p_i}{p_{n-1}} - 1\right)$ approaches zero. This suggests a great indeterminacy in investors' trading intention toward the stock under consideration [35,36].

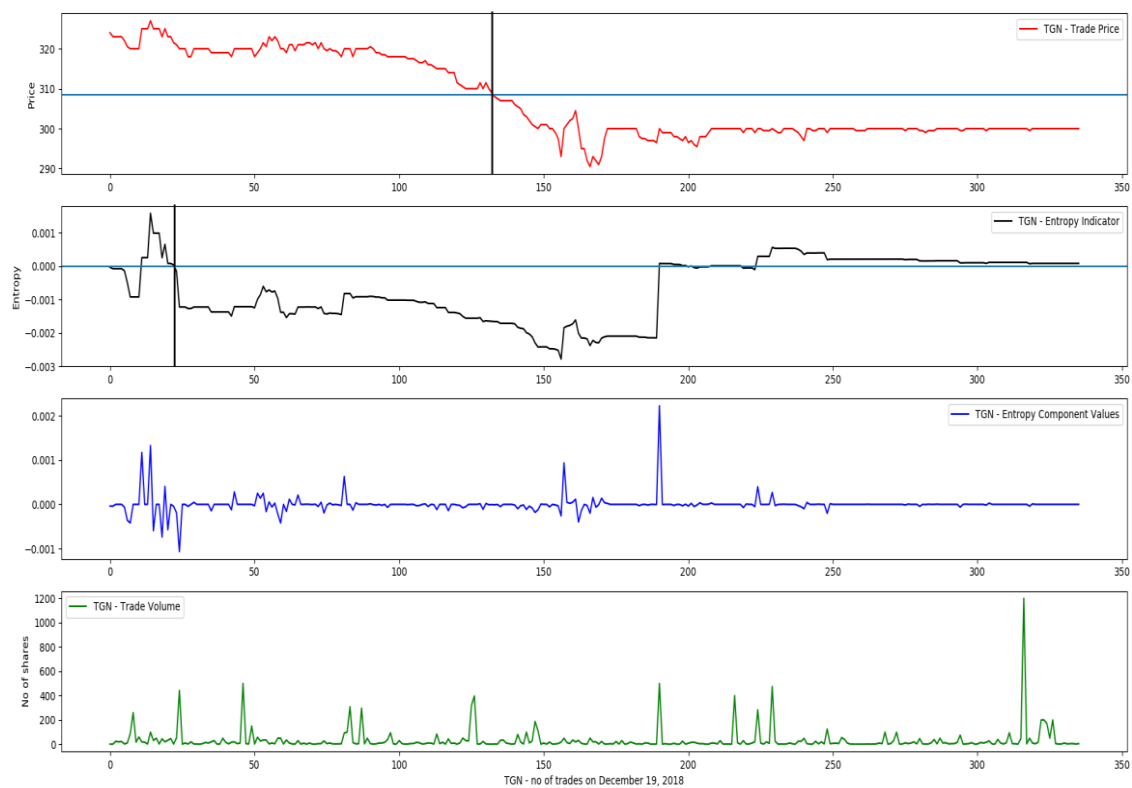

**Figure 4.** Negative entropy values signal prevailing inclination of the market to sell; later on, the stock price follows and goes under the VWAP of the day of December 19, 2018 (symbol TGN).

Figures A2 and A3 (Appendix A) provide data supplemental to the empirical results obtained by computing the intrinsic entropy for additional couple of symbols listed on the BVB. The historical production data represent transactions executed on the BVB in December 2018. We want to emphasize that the computed intrinsic entropy goes into the negative territory well in advance the moment trade price goes under the VWAP line. The continuous negative values for the intrinsic entropy signal the investors inclination to sell the considered stock, a tendency that is later confirmed by the trade prices reaching levels below VWAP.

## 4. Employing the Proposed Intrinsic Entropy in Algorithmic Trading

In order to demonstrate the usability of our intrinsic entropy model for practitioners, we designed and implemented two intraday trading algorithms:

(1) one that makes decisions based on the values of the computed entropy along with the VWAP;
(2) one that uses only the VAWP in the decision-making process.

The VWAP is widely used by investors as a moving price reference for intraday trading to estimate the yield of transaction [37]. We test both trading algorithms against historical data consisting of trades executed on the BVB throughout 19th of December, 2018. It is was day following the announcement made by the Romanian Finance Minister regarding the introduction of a "greed tax" for banks, capping the gas price for three years and lowering the commission for private pension fund administrators. These had been measures among others included in an emergency ordinance that would have been adopted by the government toward the end of the year. That particular day was a very turbulent one, and had the highest number of transactions executed in a single day on the BVB since 2008 financial crisis. Therefore, it provides a suitable data set for our purpose of having available a high number of trades executed in a single day.

Both algorithms use the same strategy: buy at a lower price than the VWAP and then sell at a higher price than the VWAP, if an opportunity arises. If the stock did not sell at a better price than the VWAP by the end of the day, then close the position at the last price of the day. Both algorithms allow for at least 10 trades to be made the exchange for the given stock, in order to compute the intrinsic entropy and the VWAP based on a reasonable number of transactions. Furthermore, the algorithms generate buy and respectively sell orders with quantity (numbers of shares) equal to 1.

For a given stock listed on the BVB, both trading algorithms determine:

(a) whether to buy the stock or not;
(b) when to buy it;
(c) when to sell by the end of the trading day.

The trading algorithm that uses the intrinsic entropy, we compute 3 variants of it:

(1) the intrinsic entropy that uses the opening price of the day as reference ($H_1$);
(2) the intrinsic entropy that uses the price of the proceeding trade as reference ($H_k$);
(3) the intrinsic entropy that uses the VWAP computed for the previous trade as reference ($H_{vwap}$).

The following pseudo-code describes the algorithm that employs the intrinsic entropy. For making the decision to buy a given symbol then at least one of the three variants of computed entropy have to be positive and the buy price has to be lower than the VWAP. If we have a buy trade for the considered symbol, then the algorithm sells it either when all the three variants of the intrinsic entropy have negative values or when the stock is traded at higher price than the VWAP.

```
Initialization: S = the set of all symbols tradable on the BVB
// S can be initialized also with a subset of symbols selected by investor
s = first symbol in S
while (s = next symbol in S) do
    N = n       // n is the no. of trades executed for symbol during the day
    buyFlag = False
    sellFlag = False
    buyTradeNo = 0
    Initialize the list of trade prices for s
    Initialize the list of moving VWAP for s
    Initialize the list of intrinsic entropy values H₁ for s
    Initialize the list of intrinsic entropy values Hₖ for s
    Initialize the list of intrinsic entropy values Hᵥwap for s
    for i=10 to N do
        if ((H₁[i] > 0) or (Hₖ[i] > 0) or (Hᵥwap[i] > 0)) and
              (p[i] < VWAP[i]) and (buyFlag is False) then
            Buy symbol s
            buyFlag = True
            buyTradeNo = i
        end if
        if (i > buyTradeNo) and (((H₁[i] < 0) and (Hₖ[i] < 0) and (Hᵥwap[i] < 0)) or
              (p[i] < VWAP[i]) or (i >= N-1)) and
              (buyFlag is True) and (sellFlag is False) then
            Sell symbol s
            sellFlag = True
        end if
    end for
end while
```

On the BVB are listed 83 companies (as of November 15 2019). For both algorithms we use a set (*S*) of 29 most traded stocks on the BVB on 19[th] of December 2018. The algorithm employing the intrinsic entropy model chose to buy and sell 11 symbols out of 29. Table A1 (Appendix A) shows all the symbols that were taken into account. The ones that were not traded did not satisfy the buy condition of having at least one of the three variants of computed entropy positive. The Table 2 summarizes only the symbols that were traded through the trading algorithm based on the intrinsic entropy model proposed in this paper.

Table 2. The list of trades created by the trading algorithm that employs the intrinsic entropy model; trade data executed on the BVB on December 19, 2018.

| No. | Symbol | Trade Side | Buy Trade No. | Buy Price | Buy VWAP | Trade Side | Sell Trade No. | Sell Price | Sell VWAP | Return (%) |
|---|---|---|---|---|---|---|---|---|---|---|
| 1 | CMP | BUY | 12 | 0.87 | 0.908 | SELL | 35 | 0.88 | 0.892 | 1.1494 |
| 2 | ALU | BUY | 30 | 0.65 | 0.67 | SELL | 33 | 0.665 | 0.67 | 2.3077 |
| 3 | BRD | BUY | 10 | 11.7 | 11.999 | SELL | 11 | 12 | 11.998 | 2.5641 |
| 4 | CEON | BUY | 10 | 0.3 | 0.301 | SELL | 15 | 0.3 | 0.299 | 0 |
| 5 | COTE | BUY | 62 | 75.2 | 76.864 | SELL | 63 | 77 | 76.869 | 2.3936 |
| 6 | SIF2 | BUY | 11 | 1.13 | 1.156 | SELL | 13 | 1.17 | 1.156 | 3.5398 |
| 7 | SNG | BUY | 95 | 32.75 | 32.77 | SELL | 508 | 31.7 | 31.482 | −3.2061 |
| 8 | SNN | BUY | 29 | 9.01 | 9.466 | SELL | 30 | 9.5 | 9.466 | 5.4384 |
| 9 | SNP | BUY | 12 | 0.32 | 0.33 | SELL | 14 | 0.33 | 0.33 | 3.125 |
| 10 | TGN | BUY | 11 | 320 | 322.692 | SELL | 12 | 325 | 322.731 | 1.5625 |
| 11 | TRP | BUY | 36 | 0.259 | 0.259 | SELL | 37 | 0.26 | 0.259 | 0.3861 |
| | | | | | **Cumulated return** | | | | | **19.2605** |

The following pseudo-code describes the algorithm that is based only on the VWAP. For making the decision to buy a given symbol then the buy price has to be lower than the VWAP. If we have a buy for the considered symbol, then the algorithm sells it when the stock is traded at higher price than the VWAP.

```
Initialization: S = the set of all symbols tradable on the BVB
// S can be initialized also with a subset of symbols selected by investor
s = first symbol in S
while (s = next symbol in S) do
    N = n        // n is the no. of trades executed for symbol s during the day
    buyFlag = False
    sellFlag = False
    buyTradeNo = 0
    Initialize the list of trade prices for s
    Initialize the list of moving VWAP for s
    for i=10 to N do
        if (p[i] < VWAP[i]) and (buyFlag is False) then
            Buy symbol s
            buyFlag = True
            buyTradeNo = i
        end if
        if (i > buyTradeNo) and ((p[i] < VWAP[i]) or (i >= N-1)) and
              (buyFlag is True) and (sellFlag is False) then
            Sell symbol s
            sellFlag = True
        end if
    end for
end while
```

The algorithm based only on the VWAP chose to buy and sell the all 29 stocks from the test data set. Table A2 (Appendix A) shows all the symbols that were taken into account. Table 3 summarizes only the symbols that were traded also by the algorithm that employs the intrinsic entropy model in order to compare the returns generated by both algorithms for the same traded symbols.

**Table 3.** The list of trades created by the trading algorithm based only on the VWAP; trade data executed on the BVB on December 19, 2018.

| No. | Symbol | Trade Side | Buy Trade No. | Buy Price | Buy VWAP | Trade Side | Sell Trade No. | Sell Price | Sell VWAP | Return (%) |
|---|---|---|---|---|---|---|---|---|---|---|
| 1 | CMP | BUY | 12 | 0.87 | 0.908 | SELL | 43 | 0.822 | 0.886 | −5.5172 |
| 2 | ALU | BUY | 10 | 0.66 | 0.68 | SELL | 19 | 0.68 | 0.676 | 3.0303 |
| 3 | BRD | BUY | 10 | 11.7 | 11.999 | SELL | 11 | 12 | 11.999 | 2.5641 |
| 4 | CEON | BUY | 10 | 0.3 | 0.301 | SELL | 15 | 0.3 | 0.299 | 0 |
| 5 | COTE | BUY | 11 | 78 | 78.5 | SELL | 63 | 77 | 76.864 | −1.2821 |
| 6 | SIF2 | BUY | 10 | 1.124 | 1.177 | SELL | 13 | 1.17 | 1.156 | 4.0925 |
| 7 | SNG | BUY | 95 | 32.75 | 32.772 | SELL | 508 | 31.7 | 31.482 | −3.2061 |
| 8 | SNN | BUY | 10 | 9.46 | 9.577 | SELL | 30 | 9.5 | 9.466 | 0.4228 |
| 9 | SNP | BUY | 10 | 0.322 | 0.33 | SELL | 14 | 0.33 | 0.33 | 2.4845 |
| 10 | TGN | BUY | 10 | 320 | 322.82 | SELL | 12 | 325 | 322.692 | 1.5625 |
| 11 | TRP | BUY | 10 | 0.256 | 0.257 | SELL | 12 | 0.258 | 0.257 | 0.7813 |
| | | | | | **Cumulated return** | | | | | **4.9326** |

The comparison between the returns generated by these two trading algorithms shows the following:

- the trading algorithm that uses only the VWAP generates a cumulated return of 4.93%, or 0.448% in average for the 11 traded symbols;
- the trading algorithm that employs the intrinsic entropy model that we propose here produces a cumulated return of 19.26%, or 1.75% in average for the 11 traded symbols.

We obtained a 3.9 times higher return by incorporating the intrinsic entropy model in an otherwise simple VWAP based trading algorithm.

When computed for each instrument listed on the BVB, the intrinsic entropy can provide a synthetic perspective of the overall market performance. In order to further assist the practitioners in selecting a specific subset of the listed stocks for a certain trading algorithm, we create the intrinsic entropy market map [38]. The map allows for timely assessing the investors' interest in the securities listed on the BVB. Details are shown in Figure A1 (Appendix A). Each rectangle on the map represents a listed company on the BVB. The size of the rectangle is proportionally determined, based on the relative weight of the individual market cap of a company in the overall market capitalization. The rectangles are coloured based on the value of intrinsic entropy computed for each stock:

- nuances of red for negative entropies;
- black for entropy levels around zero;
- nuances of green for positive entropy levels.

## 5. Conclusions and Further Research

In this paper we introduce an entropic model which can be employed for gauging the investors' interest in a given exchange-traded security, along with the state of the overall market, corroborated by the individual security trading data.

The data employed for testing consisted of historical intraday transactions executed on the BVB. The empirical test results prove that the proposed intrinsic entropy model can be used as an indicator for evaluating the direction and the intensity of intraday trading activity of an exchange-traded security. The intrinsic entropy signals if the market is either inclined to buy the security or rather to sell it. The model that we propose does not include any exogenous factor. Any external factors likely to influence the intrinsic entropy are embedded. The trading data (volume, price, trade frequency and the number of trades) already contains them, and they are quantified within the intrinsic entropy model. The intrinsic entropy signals if the market is either inclined to buy the security or rather to sell it. We further explore the usage of the intrinsic entropy model for algorithmic trading and demonstrate the value of our model in assisting investors' intraday stock portfolio selection, along with timely generated signals for supporting the buy/sell decision-making process. The data employed for testing consisted of historical intraday transactions executed on the BVB.

In order to demonstrate the usability of our intrinsic entropy model for practitioners, we designed and implemented two intraday trading algorithms: one that makes decisions based on the values of the computed entropy along with the VWAP, and another one that uses only the VAWP in the decision-making process. The comparison between the returns generated by the two trading algorithms shows obtained a return 3.9 times higher by incorporating the intrinsic entropy model in an otherwise simple VWAP based trading algorithm. In addition to these results, it has to be noticed that the trading algorithm based on the intrinsic entropy provided an embedded mechanism for stock portfolio selection: it traded 11 out of 29 considered stocks.

When computed for each instrument listed on the BVB, the intrinsic entropy can provide a synthetic perspective of the overall market performance. In order to further assist the practitioners in selecting a specific subset of the listed stocks for a certain trading algorithm, we create the intrinsic entropy market map. The map allows for timely assessing the investors' interest in the securities listed on the BVB.

We are aware that the intrinsic entropy model that we propose here may provide less support when built over consecutive trading days. Nevertheless, having given the relatively low number of transactions per day executed on the BVB for each listed stock, a longer time frame would provide more data for supporting more recent activity. For instance, even for intraday computed intrinsic entropy, it would be advantageous not to start from zero at the beginning of the trading day, but to potentially carry over some market information from the previous day. Yet, the focus should be on the most recent data, which we consider to have a greater relevance for the intrinsic entropy.

There are further usages to be investigated for this intrinsic entropy model. For example, the Black-Scholes-Merton (BSM) model estimates the variation over time of financial instruments such as stocks, and using the implied volatility of the underlying security derives the price of a call option. The intrinsic entropy model might prove to be a better anchored substitute for the implied volatility used in the BSM model, since it is computed based on the actual trading data of the underlying stock.

This paper presents the early results of our research, and we expect the scientific community to identify other new usages of the intrinsic entropy model for exchange-traded securities.

**Author Contributions:**
The authors contributed comprehensively to the paper, according each one competence, experience, interest and background.

**Funding:**
This research received no external funding.

**Conflicts of Interest**:
The authors declare no conflict of interest.

# Appendix A

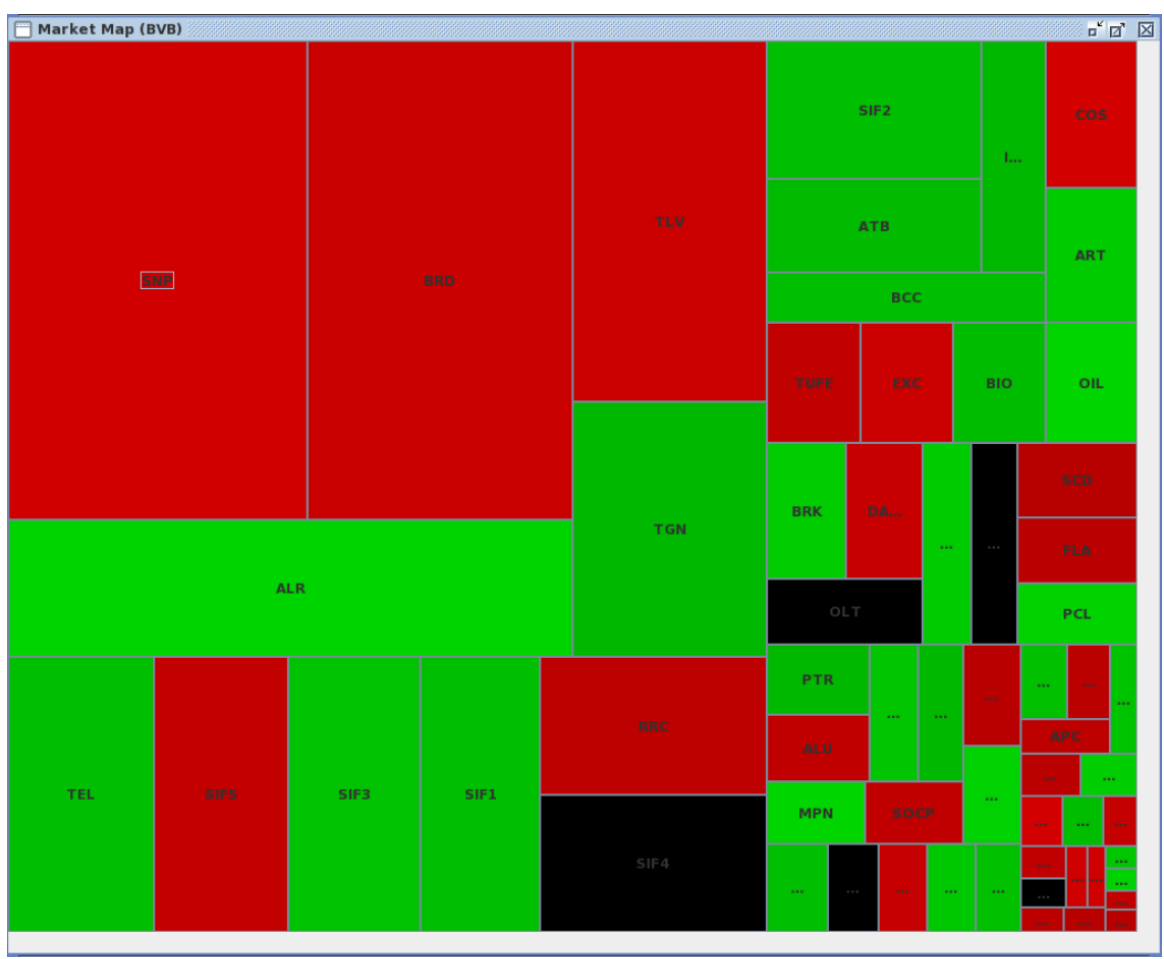

**Figure A1.** Stock Intrinsic Entropy Market Map for The Bucharest Stock Exchange (BVB) for the day of 18 December 2018.

Table A1. The list of all the symbols considered and the trades generated by the trading algorithm that employs the intrinsic entropy model; trade data executed on the BVB on December 19, 2018.

| No. | Symbol | Trade Side | Buy Trade No. | Buy Price | Buy VWAP | Trade Side | Sell Trade No. | Sell Price | Sell VWAP | Return (%) |
|---|---|---|---|---|---|---|---|---|---|---|
| 1 | CMP | BUY | 12 | 0.87 | 0.908 | SELL | 35 | 0.88 | 0.892 | 1.1494 |
| 2 | ALR | | | | | | | | | |
| 3 | ALU | BUY | 30 | 0.65 | 0.67 | SELL | 33 | 0.665 | 0.67 | 2.3077 |
| 4 | ATB | | | | | | | | | |
| 5 | BRD | BUY | 10 | 11.7 | 11.999 | SELL | 11 | 12 | 11.998 | 2.5641 |
| 6 | BVB | | | | | | | | | |
| 7 | CEON | BUY | 10 | 0.3 | 0.301 | SELL | 15 | 0.3 | 0.299 | 0 |
| 8 | COTE | BUY | 62 | 75.2 | 76.864 | SELL | 63 | 77 | 76.869 | 2.3936 |
| 9 | EBS | | | | | | | | | |
| 10 | DIGI | | | | | | | | | |
| 11 | EL | | | | | | | | | |
| 12 | ELMA | | | | | | | | | |
| 13 | FP | | | | | | | | | |
| 14 | M | | | | | | | | | |
| 15 | PBK | | | | | | | | | |
| 16 | ROCE | | | | | | | | | |
| 17 | SIF1 | | | | | | | | | |
| 18 | SIF2 | BUY | 11 | 1.13 | 1.156 | SELL | 13 | 1.17 | 1.156 | 3.5398 |
| 19 | SIF3 | | | | | | | | | |
| 20 | SIF4 | | | | | | | | | |
| 21 | SIF5 | | | | | | | | | |
| 22 | SNG | BUY | 95 | 32.75 | 32.77 | SELL | 508 | 31.7 | 31.482 | −3.2061 |
| 23 | SNN | BUY | 29 | 9.01 | 9.466 | SELL | 30 | 9.5 | 9.466 | 5.4384 |
| 24 | SNP | BUY | 12 | 0.32 | 0.33 | SELL | 14 | 0.33 | 0.33 | 3.125 |
| 25 | TBM | | | | | | | | | |
| 26 | TEL | | | | | | | | | |
| 27 | TGN | BUY | 11 | 320 | 322.692 | SELL | 12 | 325 | 322.731 | 1.5625 |
| 28 | TLV | | | | | | | | | |
| 29 | TRP | BUY | 36 | 0.259 | 0.259 | SELL | 37 | 0.26 | 0.259 | 0.3861 |

**Table A2.** The list of all the symbols considered and the trades generated by the trading algorithm based only on the VWAP; trade data executed on the BVB on December 19, 2018.

| No. | Symbol | Trade Side | Buy Trade No. | Buy Price | Buy VWAP | Trade Side | Sell Trade No. | Sell Price | Sell VWAP | Return (%) |
|---|---|---|---|---|---|---|---|---|---|---|
| 1 | CMP | BUY | 12 | 0.87 | 0.908 | SELL | 43 | 0.822 | 0.886 | −5.5172 |
| 2 | ALR | BUY | 10 | 3.28 | 3.351 | SELL | 26 | 3.28 | 3.267 | 0 |
| 3 | ALU | BUY | 10 | 0.66 | 0.68 | SELL | 19 | 0.68 | 0.676 | 3.0303 |
| 4 | ATB | BUY | 10 | 0.479 | 0.481 | SELL | 14 | 0.486 | 0.48 | 1.4614 |
| 5 | BRD | BUY | 10 | 11.7 | 11.999 | SELL | 11 | 12 | 11.999 | 2.5641 |
| 6 | BVB | BUY | 10 | 22.1 | 22.273 | SELL | 18 | 22.2 | 22.192 | 0.4525 |
| 7 | CEON | BUY | 10 | 0.3 | 0.301 | SELL | 15 | 0.3 | 0.299 | 0 |
| 8 | COTE | BUY | 11 | 78 | 78.5 | SELL | 63 | 77 | 76.864 | −1.2821 |
| 9 | EBS | BUY | 10 | 139.05 | 141.371 | SELL | 29 | 149 | 139.265 | 7.1557 |
| 10 | DIGI | BUY | 10 | 25.6 | 26.295 | SELL | 21 | 26 | 25.84 | 1.5625 |
| 11 | EL | BUY | 11 | 10.7 | 10.81 | SELL | 265 | 9.98 | 10.045 | −6.729 |
| 12 | ELMA | BUY | 10 | 0.141 | 0.143 | SELL | 21 | 0.146 | 0.141 | 3.5461 |
| 13 | FP | BUY | 10 | 0.92 | 0.923 | SELL | 87 | 0.914 | 0.911 | −0.6522 |
| 14 | M | BUY | 10 | 26.9 | 27.005 | SELL | 13 | 27 | 26.899 | 0.3717 |
| 15 | PBK | BUY | 10 | 0.071 | 0.075 | SELL | 21 | 0.074 | 0.072 | 4.8159 |
| 16 | ROCE | BUY | 10 | 0.11 | 0.113 | SELL | 15 | 0.118 | 0.113 | 7.2727 |
| 17 | SIF1 | BUY | 10 | 2.2 | 2.216 | SELL | 11 | 2.26 | 2.216 | 2.7273 |
| 18 | SIF2 | BUY | 10 | 1.124 | 1.177 | SELL | 13 | 1.17 | 1.156 | 4.0925 |
| 19 | SIF3 | BUY | 10 | 0.207 | 0.209 | SELL | 24 | 0.207 | 0.207 | 0 |
| 20 | SIF4 | BUY | 10 | 0.6 | 0.603 | SELL | 128 | 0.574 | 0.58 | −4.3333 |
| 21 | SIF5 | BUY | 10 | 2.05 | 2.061 | SELL | 44 | 2.05 | 2.044 | 0 |
| 22 | SNG | BUY | 95 | 32.75 | 32.772 | SELL | 508 | 31.7 | 31.482 | −3.2061 |
| 23 | SNN | BUY | 10 | 9.46 | 9.577 | SELL | 30 | 9.5 | 9.466 | 0.4228 |
| 24 | SNP | BUY | 10 | 0.322 | 0.33 | SELL | 14 | 0.33 | 0.33 | 2.4845 |
| 25 | TBM | BUY | 10 | 0.205 | 0.214 | SELL | 15 | 0.22 | 0.213 | 7.3171 |
| 26 | TEL | BUY | 10 | 22.25 | 22.373 | SELL | 51 | 22.2 | 22.111 | −0.2247 |
| 27 | TGN | BUY | 10 | 320 | 322.82 | SELL | 12 | 325 | 322.692 | 1.5625 |
| 28 | TLV | BUY | 10 | 2.045 | 2.05 | SELL | 146 | 2.04 | 2.026 | −0.2445 |
| 29 | TRP | BUY | 10 | 0.256 | 0.257 | SELL | 12 | 0.258 | 0.257 | 0.7813 |

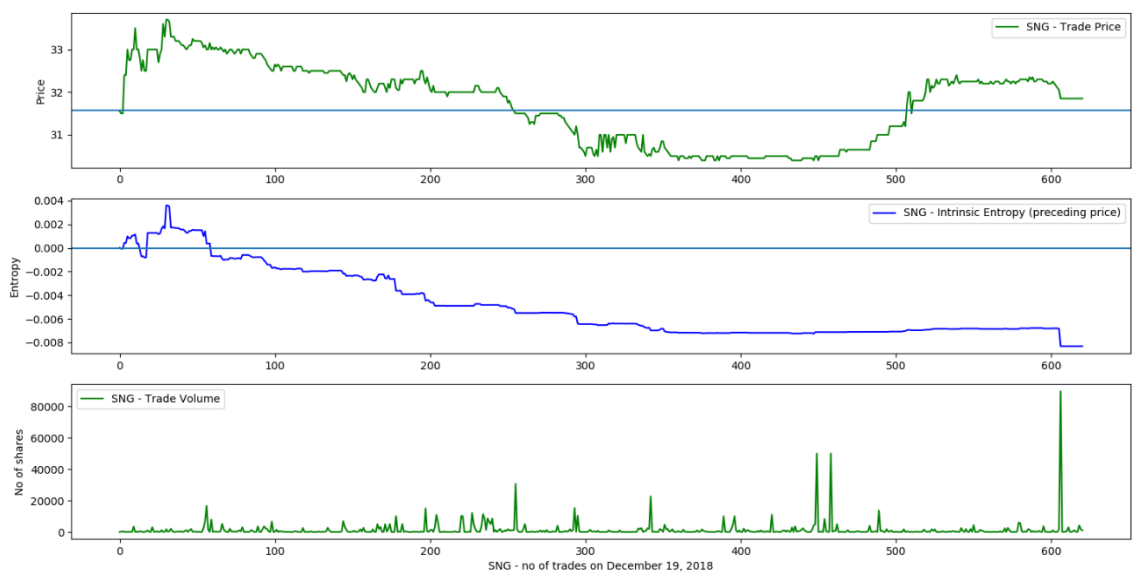

**Figure A2.** Price variation, entropy, entropy component values and trade volumes for the day of December 19, 2018, symbol SNG (Romgaz).

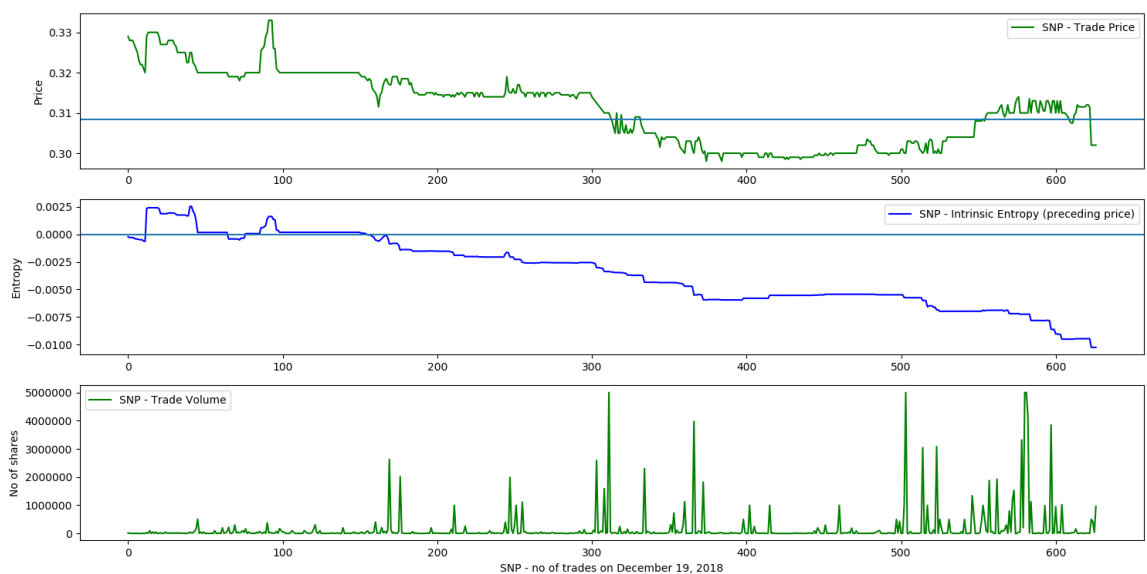

**Figure A3.** Price variation, entropy, entropy component values and trade volumes for the day of December 19, 2018, symbol SNP (OMV Petrom).

## References

1. Duan, W. The Sales Impact of Word-of-Mouth Distribution across Retail and Third-Party Websites. In Proceedings of the Sales Impact of the Distribution of Online WOM Thirty Seventh International Conference on Information Systems, Dublin, 11–14 December 2016.
2. Sprenger, T.O. and Welpe, I.M. News or Noise? The Stock Market Reaction to Different Types of Company-Specific News Events (January 4, 2011). Available at SSRN: https://ssrn.com/abstract=1734632 or http://dx.doi.org/10.2139/ssrn.1734632
3. Machado, J.A.; Duarte, F.B.; Duarte, G.M. Analysis of stock market indices with multidimensional scaling and wavelets. *Math. Probl. Eng.* **2012**, *2012*, 1–14.
4. Fama, E.F. Efficient Capital Markets: A Review of Theory and Empirical Work. *J. Financ.* **1970**, *25*, 383–417.
5. Frigg, R.; Werndl, C. Entropy—A Guide for the Perplexed. In *Probabilities in Physics*; Beisbart, C., Hartmann, S., Eds.; Oxford University Press: Oxford, UK, 2010.
6. Shannon, E.C. A Mathematical Theory of Communication. *Bell Syst. Tech. J.* **1948**, *27*, 379–423.
7. Clausius, R. On a Modified Form of the Second Fundamental Theorem in the Mechanical Theory of Heat. *Lond. Edinb. Dublin Philos. Mag. J. Sci.* **1856**, *12*, 81–98.
8. Boltzmann, L. *Lectures on Gas Theory*; translated by S.G.; Brush, University of California Press: Berkeley, CS, USA, 2012.
9. Schneider, T.D. *Information Theory Primer with an Appendix on Logarithms*; National Cancer Institute: Bethesda, MD, USA, 2007; pp. 5–6.
10. Jaynes, E.T.; Smith, C.R.; Erickson, G.J. The Gibbs Paradox. In *Maximum Entropy and Bayesian Methods*; Neudorfer, P.O., Ed.; Kluwer Academic: Dordrecht, The Netherlands, 1992; pp. 1–22.
11. Georgescu-Roegen, N. *The Entropy Law and the Economic Process*; Harvard University Press: Cambridge, MA, USA, 1971.
12. Ausloos, M. New region planning in France? Better order or more disorder? *Entropy* **2015**, *17*, 5695–5710.
13. Ausloos, M.; Nedic, O.; Dekanski, A. Seasonal Entropy, Diversity and Inequality Measures of Submitted and Accepted Papers Distributions in Peer-Reviewed Journals. *Entropy* **2019**, *21*, 564.
14. Marti, G.; Nielsen, F.; Bińkowski, M.; Donnat, P. A review of two decades of correlations, hierarchies, networks and clustering in financial markets. *arXiv Prepr.* **2017**, arXiv:1703.00485.
15. Horowitz, R.A.; Horowitz, I. The real and illusory virtues of entropy-based measures for business and economic analysis. *Decis. Sci.* **1976**, *7*, 121–136.
16. Philippatos, G.; Wilson, C. Entropy, market risk and the selection of efficient portfolios. *Appl. Econ.* **1972**, *4*, 209–220.
17. Philippatos, G.; Wilson, C. Entropy, market risk and the selection of efficient portfolios: Reply. *Appl. Econ.* **1974**, *6*, 76–79.
18. White, D. Entropy, market risk and the selection of efficient portfolios: Comment. *Appl. Econ.* **1974**, *6*, 73–75.
19. Nawrocki, D.N.; Harding, W.H. State-value weighted entropy as a measure of investment risk. *Appl. Econ.* **1986**, *18*, 411–419.
20. Tsallis, C. Possible generalization of Boltzmann–Gibbs statistics. *J. Stat. Phys.* **1988**, *52*, 479–487.
21. Jaynes, E.T. Gibbs vs Boltzmann entropies. *Am. J. Phys.* **1965**, *33*, 391–398.
22. Gibbs, J.W. On the Equilibrium of Heterogeneous Substances. *Trans. Conn. Acad. Arts Sci.* **1878**, *16*, 441–458.
23. Gibbs, J.W. *Elementary Principles in Statistical Mechanics, Developed with Especial Reference to the Rational Foundation of Thermodynamics*; Charles Scribner's Sons: New York, NY, USA, 1902.
24. Maasoumi, E.; Racine, J. Entropy and predictability of stock market returns. *J. Econom.* **2002**, *107*, 291–312.
25. Ausloos, M.; Ivanova, K. Dynamical model and nonextensive statistical mechanics of a market index on large time windows. *Phys. Rev. E* **2003**, *68*, 046122.
26. Zhou, R.; Cai, R.; Tong, G. Applications of entropy in finance: A review. *Entropy* **2013**, *15*, 4909–4931.
27. Xu, M.; Shang, P.; Huang, J. Modified generalized sample entropy and surrogate data analysis for stock markets. *Commun. Nonlinear Sci. Numer. Simul.* **2016**, *35*, 17–24.
28. Zunino, L.; Zanin, M.; Tabak, B.M.; Pérez, D.G.; Rosso, O.A. Complexity-entropy causality plane: A useful approach to quantify the stock market inefficiency. *Phys. A Stat. Mech. Its Appl.* **2010**, *389*, 1891–1901.
29. Zanin, M.; Zunino, L.; Rosso, O.A.; Papo, D. Permutation entropy its main biomedical econophysics applications: A review. *Entropy* **2012**, *14*, 1553–157.

30. Li, J.; Liang, C.; Zhu, X.; Sun, X.; Wu, D. Risk contagion in Chinese banking industry: A Transfer Entropy-based analysis. *Entropy* **2013**, *15*, 5549–5564.
31. Cerqueti, R.; Rotundo, G.; Ausloos, M. Investigating the configurations in cross-shareholding: A joint copula-entropy approach. *Entropy* **2018**, *20*, 134.
32. Ceptureanu, S.I.; Ceptureanu, E.G.; Marin, I. Assessing the Role of Strategic Choice on Organizational Performance by Jacquemin–Berry Entropy Index. *Entropy* **2017**, *19*, 448.
33. Vinte, C. The Informatics of the Equity Markets—A Collaborative Approach. *Econ. Inform.* **2009**, *13*, 76–85.
34. Vinte, C. Methods and Apparatus for Optimizing the Distribution of Trading Executions. U.S. Patent 2014/0149274A1, 29 May 2014. Available online: https://patents.google.com/patent/US20140149274 (accessed on 22 November 2019).
35. Hoskisson, R.E.; Hitt, M.A.; Johnson, R.A.; Moesel, D.D. Construct validity of an objective (entropy) categorical measure of diversification strategy. *Strateg. Manag. J.* **1993**, *14*, 215–235.
36. Bera, A.K.; Park, S.Y. Optimal portfolio diversification using the maximum entropy principle. *Econom. Rev.* **2008**, *27*, 484–512.
37. Barry, J. *Algorithmic Trading & DMA an Introduction to Direct Access Trading Strategies*; 4Myeloma Press: London, UK, 2010; pp. 115–160.
38. Vinte, C. Upon a Tridimensional Perspective of the Stock Market. In Proceedings of the Ninth International Conference on Informatics in Economy, Bucharest, Romania, 7–8 May 2009.